\documentclass[twocolumn,showpacs,amsmath,amssymb,prl]{revtex4}

\usepackage{graphicx}% Include figure files
\usepackage{dcolumn}% Align table columns on decimal point
\usepackage{bm}% bold math

\begin{document}
%\preprint{CERC}
\title{
Role of relaxation in spin Hall effect
}
% Force line breaks with \\
%
\author{Masaru Onoda$^{1,3}$}
\email{m.onoda@aist.go.jp}
\author{Naoto Nagaosa$^{1,2,3}$}
\email{nagaosa@appi.t.u-tokyo.ac.jp}
\affiliation{
$^1$Correlated Electron Research Center (CERC),
National Institute of Advanced Industrial Science and Technology (AIST),
Tsukuba Central 4, Tsukuba 305-8562, Japan\\
$^2$Department of Applied Physics, University of Tokyo,
Bunkyo-ku, Tokyo 113-8656, Japan
$^3$CREST, Japan Science and Technology Corporation (JST),
Saitama, 332-0012, Japan
}

\begin{abstract}
The role of the relaxation due to the impurity scattering and/or the 
contacts to leads/electrodes are studied for the spin Hall effect (SHE).
Relaxation is essential to attain the steady state and also to the spin 
accumulation, but has been considered to be harmful for the intrinsic 
SHE (ISHE). These issues are examined quantitatively 
on two types of 2D models, i.e., (a) Rashba model for $n$-type GaAs,
and (b) Luttinger model for $p$-type GaAs.
It is found that ISHE is robust against the realistic strength of the disorder
producing the observable amount of the spin accumulation. Especially in model 
(b) the spin current and the accumulation are 
an order of magnitude larger than those in model (a).
Experimental observations are discussed quantitatively from these results.
\end{abstract}
% % Classification Scheme.
%%\keywords{Suggested keywords}%Use showkeys class option if keyword
%  %display desired
\maketitle
The spin Hall effect (SHE) is a new realm of spintronics, 
by which spin current is produced perpendicular to the applied electric field. 
This enables the spin injection to the semiconductors without 
the magnets or magnetic field. This possibility has been proposed 
long ago~\cite{Dyakonov,Hirsch,Zhang}, 
but recent intensive interest comes from the theoretical proposal that 
the intrinsic mechanism due to the topological nature of the Bloch 
wavefunction in the presence of the spin-orbit interaction (SOI) 
gives rise to orders of magnitude larger effect in conventional semiconductors 
such as GaAs~\cite{MNZ,MNZ1,Sinova}. Recent two papers~\cite{Kato,Wunderlich}
reported on the experimental observation of SHE. 
One is on $n$-type GaAs where the charge current $J$ produces
the spin accumulation detected by Kerr rotation spectroscopy
near the edges of the sample transverse to $J$~\cite{Kato}. 
These authors concluded that the effect is due to the extrinsic 
origin since the effect is rather independent of the orientation of the sample.
The other is on $p$-type GaAs where also the spin accumulation is detected by 
the circularly polarized LED~\cite{Wunderlich}. 
Their sample has more carrier density compared 
with above, and the estimation of the relaxation is small which lead the 
authors to conclude the intrinsic origin.  
However the debates on the origin of the SHE, i.e.,
the intrinsic or extrinsic, continues. 
This situation is in parallel to the anomalous Hall effect (AHE), 
where the long standing controversy between the extrinsic impurity 
induced mechanism (such as the skew scattering~\cite{Smit,Kondo} 
and side jump model~\cite{Berger}) and 
intrinsic one~\cite{KL,Onoda,Jungwirth,Ong} still continues. 
Therefore it is of vital importance to study the effect of impurity 
scatterings and/or relaxation on the SHE {\it quantitatively} taking 
into account the realistic values of parameters and experimental setups, 
which we undertake in this paper.
Actually there are several preceding works addressing the issue of relaxation 
in SHE~\cite{SL,Inoue,Rashba,Raimondi,Murakami,
Hankiewitcz,Nomura,Ma,Mishchenko,Nikolic}. 
However many of the works focus on the limiting case of 
weak disorder~\cite{SL,Inoue,Rashba,Raimondi,Murakami,Ma,Mishchenko}, or 
non-equilibrium state is not
taken into account~\cite{Hankiewitcz,Nomura}, 
or lacking a quantitative comparison with existing experiments~\cite{Nikolic}.
Also the studies on the 4-band model which describes $p$-type GaAs 
are missing except ref.~\cite{Murakami}. This is partly due
to the fact that the spin relaxation is so rapid for $p$-holes since the SOI
is much larger at the top of the valence bands compared with the 
bottom of the conduction bands, which has been assumed to be disadvantageous 
for the spin accumulation. However, the strong SOI promotes the spin current
and in any case the relaxation is needed to produce the spin accumulation,
which is time reversal ($T$)-odd, from the $T$-even spin current. Therefore 
it is a nontrivial issue which is more advantageous, $p$-type or $n$-type GaAs.

 To answer all these questions, we study in this paper the 
Rashba and Luttinger models defined on the square lattice 
in terms of the Keldysh formalism 
applied to the finite size sample attached to 
the leads/electrodes~\cite{Datta,SHI-Keldysh}. 
These models reproduce the continuum version of each system
near the $\Gamma$-point of Brillouin zone, which is most relevant 
for the low carrier density. 
The energy unit $t$ (hopping parameter) and the length unit 
$a$ (lattice constant)
will be fixed later when we compare our results with experimental data.
 The model for the Rashba system is expressed as~\cite{Ando}
\begin{eqnarray}
H 
&=& \sum_{\bm{r},\bm{r}'} 
c^{\dagger}_{\bm{r}}t_{\bm{r}\bm{r}'}c_{\bm{r}'},
\\
t_{\bm{r}\bm{r}'} 
&=& \left\{
\begin{array}{rl}
-\sqrt{1-S^2}t\mp iSt\sigma_{y}, & \bm{r} = \bm{r}' \pm a\bm{e}_{x}\\
-\sqrt{1-S^2}t\pm iSt\sigma_{x}, & \bm{r} = \bm{r}' \pm a\bm{e}_{y}\\
\end{array}
\right.
\end{eqnarray}
When the Fermi energy is near the band bottom and $S\ll 1$,
the effective mass $m^{*}$ and the Rashba coupling $\alpha$
are given by $m^{*} \sim 1/(2ta^{2})$ and $\alpha \sim Sta$, respectively.

The model for the Luttinger system is defined by
\begin{eqnarray}
H 
&=& \sum_{\bm{r},\bm{r}'}\sum_{\mu=0}^{5} 
c^{\dagger}_{\bm{r}}t^{\mu}_{\bm{r}\bm{r}'}\Gamma_{\mu}c_{\bm{r}'},
\\
t^{0}_{\bm{r}\bm{r}'} 
&=&
\left\{
\begin{array}{rl}
\sqrt{1-S^{2}}t, & \bm{r} = \bm{r}' \pm a \bm{e}_{x,y}\\
-2\sqrt{1-S^{2}}t, & \bm{r} = \bm{r}'
\end{array}
\right.
,\\
t^{1,2}_{\bm{r}\bm{r}'} 
&=& 0
,\\
t^{3}_{\bm{r}\bm{r}'} 
&=& \left\{
\begin{array}{rl}
-\frac{\sqrt{3}S t}{2},& \bm{r} = \bm{r}' \pm (a \bm{e}_{x} + a \bm{e}_{y})\\
\frac{\sqrt{3}S t}{2},& \bm{r} = \bm{r}' \pm (a \bm{e}_{x} - a \bm{e}_{y})
\end{array}
\right.
,\\
t^{4}_{\bm{r}\bm{r}'} 
&=& \left\{
\begin{array}{rl}
-\sqrt{3} S t, & \bm{r} = \bm{r}' \pm a \bm{e}_{x}\\
\sqrt{3}S t, & \bm{r} = \bm{r}' \pm a \bm{e}_{y}
\end{array}
\right.
,\\
t^{5}_{\bm{r}\bm{r}'} 
&=& \left\{
\begin{array}{rl}
-S t, & \bm{r} = \bm{r}' \pm a \bm{e}_{x,y}\\
S t, & \bm{r} = \bm{r}' \pm (a \bm{e}_{x} \pm a \bm{e}_{y})\\
M, & \bm{r} = \bm{r}'
\end{array}
\right.
,
\end{eqnarray}
where $\Gamma_{0}$ is the $4\times 4$ unit matrix and 
other $\Gamma$-matrices are defined in ref.~\cite{MNZ1}. 
Assuming $p$-type GaAs thin layer, we shall take $S=0.29$, $M = 2t$.
This parameter set is corresponding to 
the typical Luttinger parameters 
$\gamma_{1} : \gamma_{2} : \gamma_{3} = 6.92 : 2.1 : 2.1$ for GaAs
and $\langle(k_{z}a)^2\rangle = M/(2St)\sim 1.86$,
is determined by the profile of the confined wavefunction along the 
$z$-direction~\cite{Bernevig}.
In the original three-dimensional system,
the kinetic terms contains $\Gamma_{1,2}$-matrices
and $t^{1,2}_{\bm{r}\bm{r}'} \neq 0$.
We can approximately neglect these terms in the quasi-two-dimensional 
system confined in a thin layer~\cite{Bernevig}.
When $t^{1,2}_{\bm{r}\bm{r}'}  = 0$, the $4\times 4$ $\Gamma$-matrix space
will be decoupled to two of the $2\times 2$ matrix space.
However, as long as we consider the doped system,
this point does not lead to any crucial difference.
We shall take the units in which $\hbar = c = 1$.

For each model, we obtain the Keldysh matrix Green function 
by solving the integral equations numerically in the self-consistent
Born approximation \cite{Datta,SHI-Keldysh}. 
The retarded self-energy is given by 
$
\Sigma^{R}_{[\bm{r}\sigma][\bm{r}'\sigma']}(E) 
= \Sigma^{R,\mathrm{cont}}_{[\bm{r}\sigma][\bm{r}'\sigma']}(E)
+\frac{-i}{2\tau_{\bm{r}}(E)}\delta_{[\bm{r}\sigma][\bm{r}'\sigma']},
$
where
$\Sigma^{R,\mathrm{cont}}(E)$ is the contact self-energy,
and $\tau_{\bm{r}}(E)$ the local lifetime due to disorders.
The local lifetime $\tau_{\bm{r}}(E)$ is determined self-consistently 
by the recursion equation,
$
\frac{1}{\tau_{\bm{r}}(E)} = \gamma N_{\bm{r}}(E)
$,
where $\gamma$ represents the strength of disorder
and $N_{\bm{r}}(E)=\frac{i}{(2\pi)}
\mathrm{Tr}^{(\sigma)}[G^{R}_{\bm{r}\bm{r}}(E)-G^{A}_{\bm{r}\bm{r}}(E)]$ 
is the local density of states per unit cell.
The lesser Green function $G^{<}_{\bm{r}\bm{r}}(E)$ is also determined 
self-consistently, by which the spatial-dependent physical 
quantities, i.e, the spin density/current, and charge density/current,
can be calculated.
We take the sample of finite size $L_{x}\times L_{y}$ with 
the electrodes attached at $x = \pm L_{x}/2$, while 
the open boundary condition is imposed in the 
$y$-direction~\cite{SHI-Keldysh}.
It is noted here that the edge modes do not play crucial role 
in the doped case.
We take the small chemical potential difference 
$\delta\mu/L_{x} = 5\times10^{-4}t/a$ to study the linear response regime.
The chemical potential in equilibrium is taken as 
$\mu_{0} = -3.5t$ for the Rashba system and
$\mu_{0} = 0$ for the Luttinger system.
Then, the Fermi energy $|\xi_{F}|$ measured from the band edge is 
$|\xi_{F}| \sim 0.5 t$ for the Rashba system and 
$|\xi_{F}| \sim 2t$ for the Luttinger system respectively.
In both systems, the dispersion near the Fermi level is
almost quadratic and isotropic.
In the Luttiger system,
the Fermi level is crossing only the heavy hole bands.
It is noted that our models describes
only the electronic states near the $\Gamma$-point
in Brillouin zone but not those of whole Brillouin zone.
Therefore, the above values of $|\xi_{F}|$
do not necessarily mean large carrier concentrations.
The substantial amount of carrier concentration 
is determined after fixing the parameters $t$ and $a$,
which will be done when we compare our results with
experimental data.

\begin{figure}[hbt]
\includegraphics[scale=0.18]{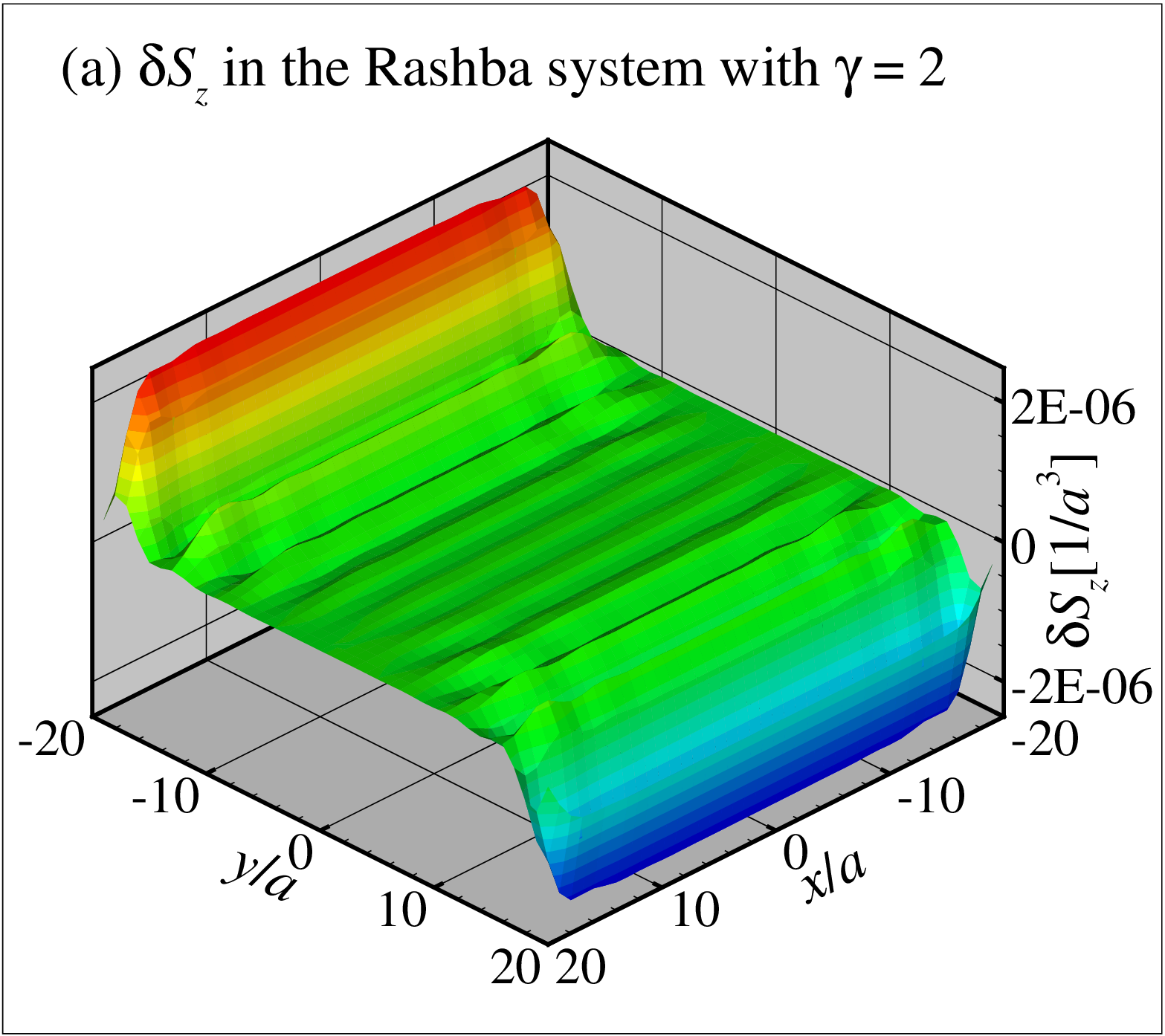}
\includegraphics[scale=0.18]{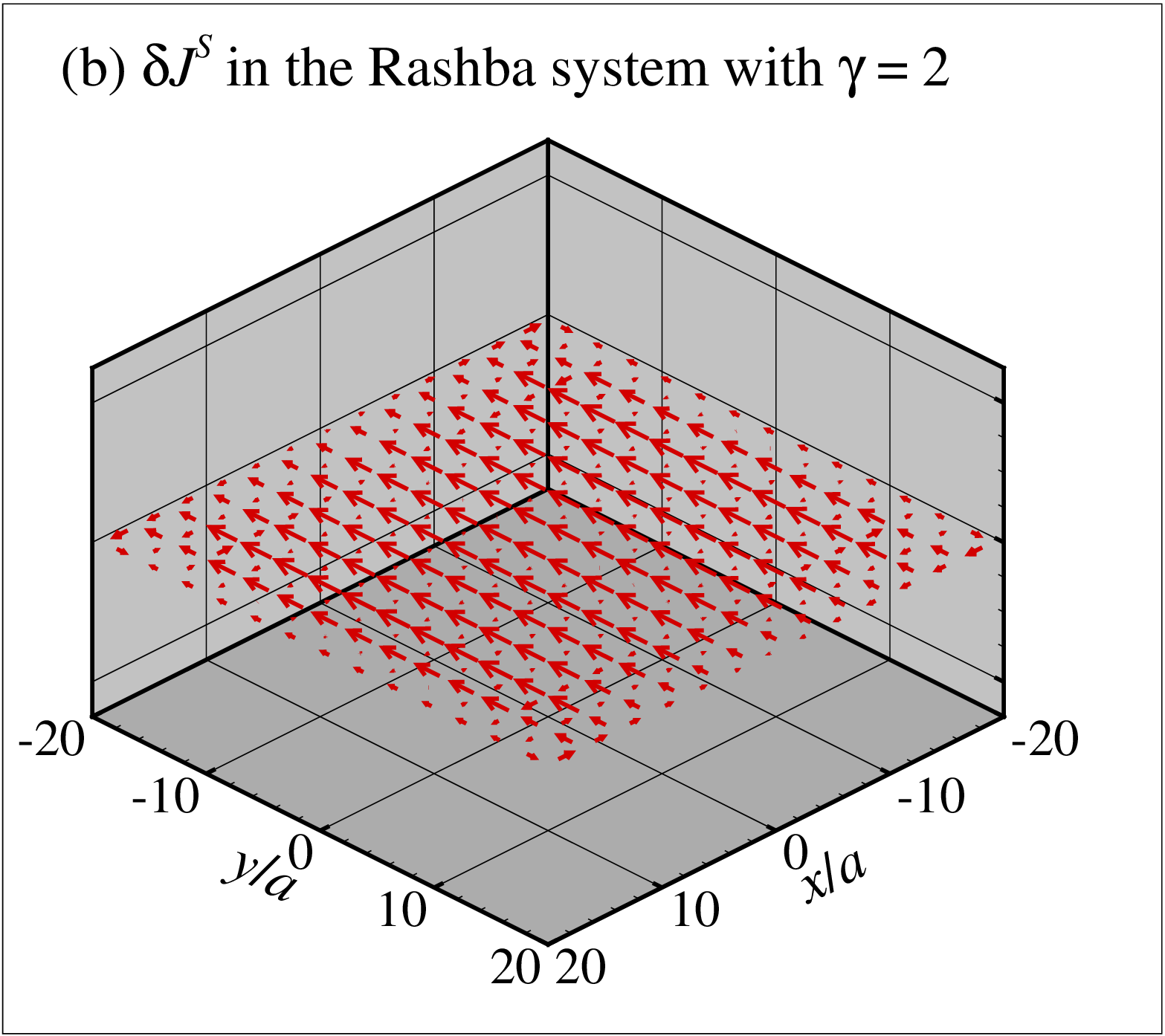}
\includegraphics[scale=0.18]{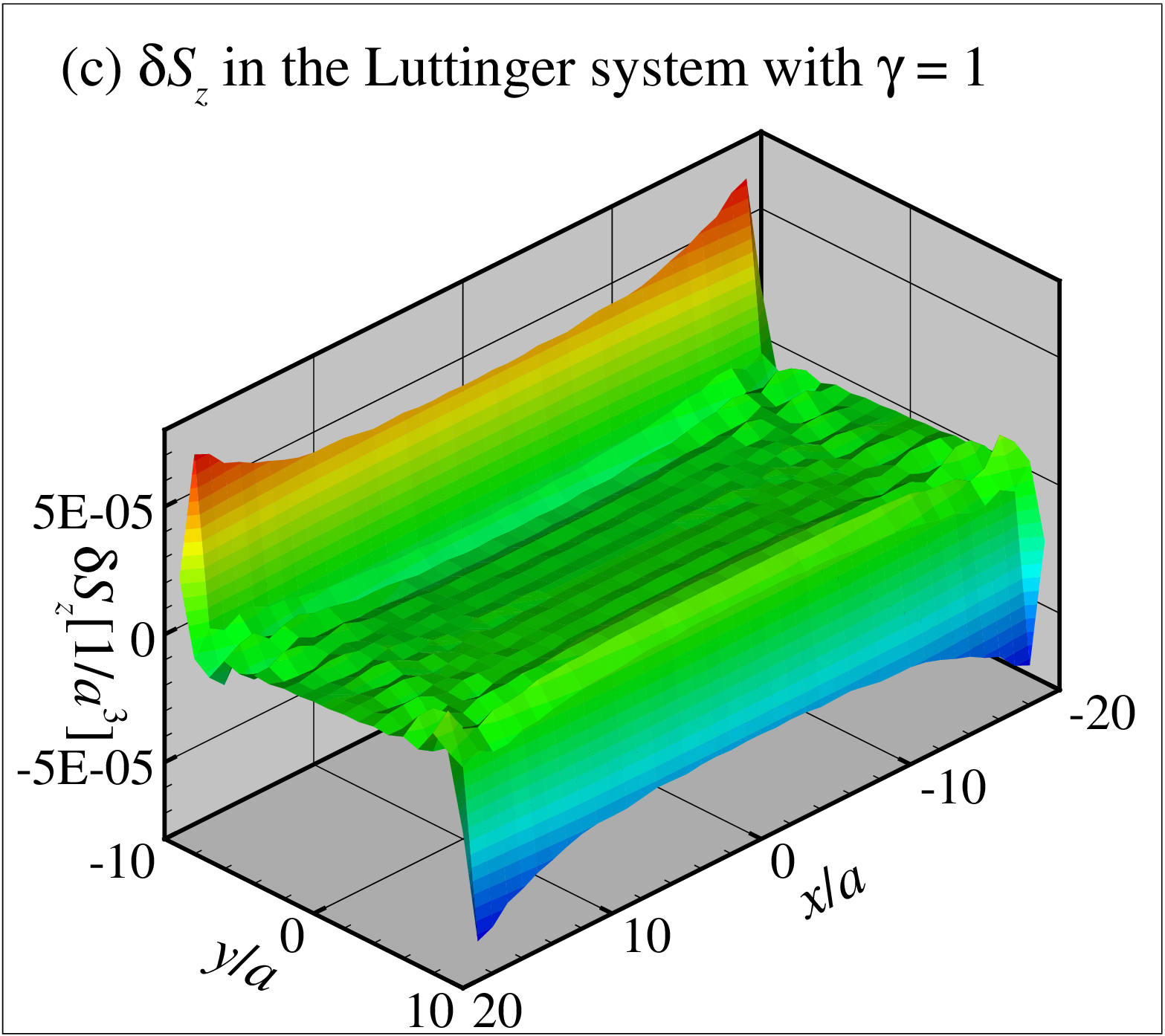}
\includegraphics[scale=0.18]{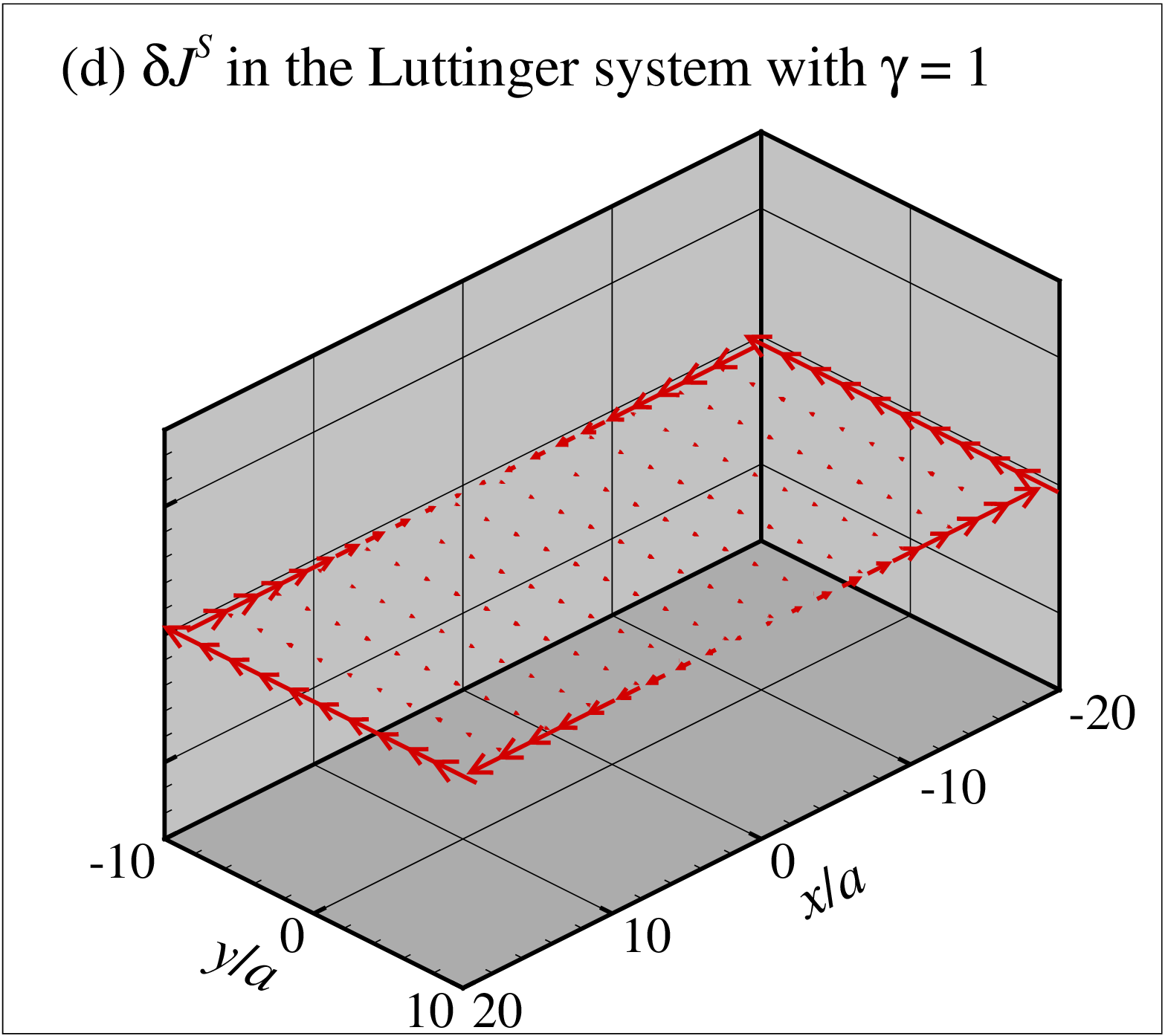}
\includegraphics[scale=0.18]{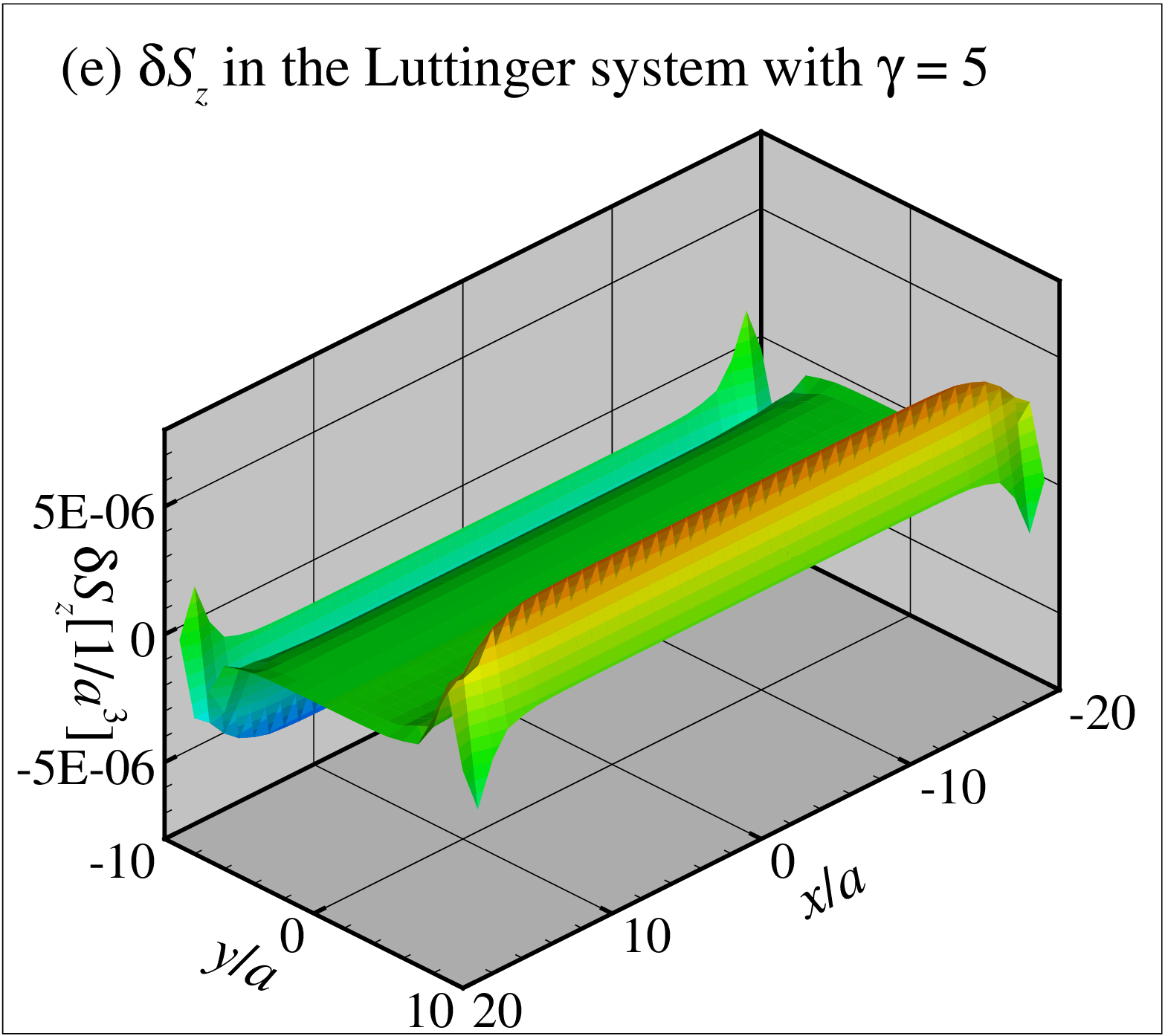}
\includegraphics[scale=0.18]{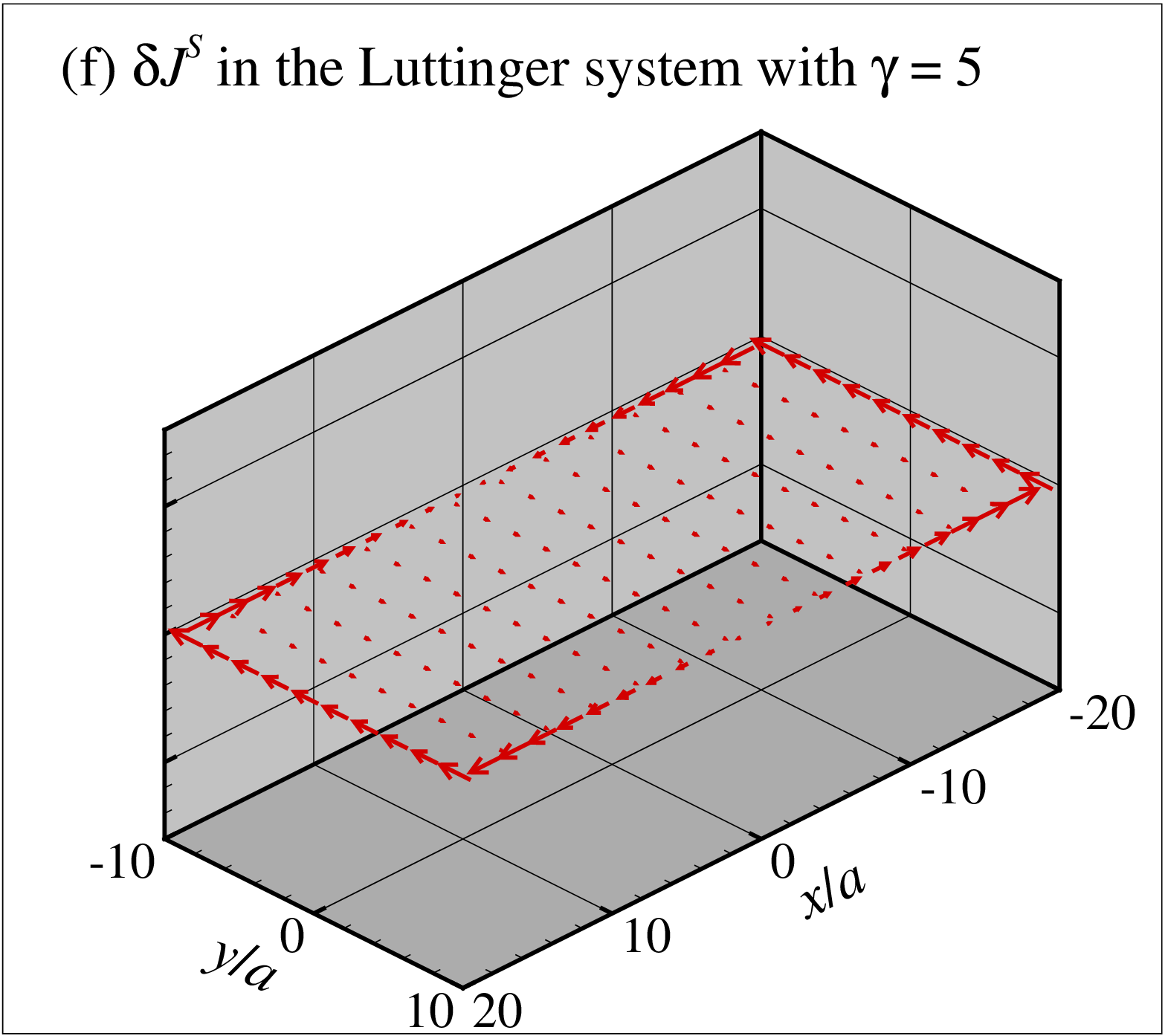}
\caption{
Spin accumulation $\delta S_z$ and 
the spin current $\delta\bm{J}^{S_{z}}$
in (a), (b) the Rashba system (S=0.05) with $\gamma = 2$ ($1/\tau \sim 0.3t$),
and the Luttinger system with (c), (d) $\gamma = 1$ ($1/\tau \sim 0.25t$) 
and (e), (f) $\gamma=5$ ($1/\tau \sim 1.25t$).
The system size is $L_{x}\times L_{y}= 40a\times40a$ for the Rashba system 
and $40a\times 20a$ for the Luttinger system.
The chemical potential of electrons is $\mu_{0}+\delta\mu/2$ 
at $x=-L_{x}/2$ and $\mu_{0}-\delta\mu/2$ at $x=L_{x}/2$. 
The charge current $\delta\bm{J}$ (not shown) flows
in the negative $x$-direction.
} 
\label{fig:dsz-dJs}
\end{figure}

\begin{figure}[hbt]
\includegraphics[scale=0.3]{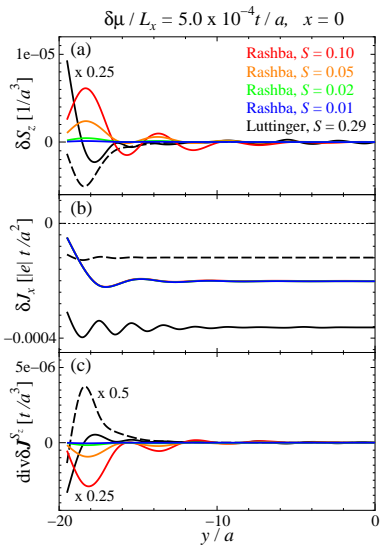}
\caption{
Distribution of (a) $\delta S_z$, (b) $\delta J_x$ 
and (c) $\bm{\nabla}\cdot\delta\bm{J}^{S_{z}}$ at $x = 0$
in the Rashba and Luttinger systems.
The system size is $L_{x}\times L_{y}= 40a\times40a$ for the Rashba system 
and $20a\times 40a$ for the Luttinger system.
Only the region $y < 0$ is shown.
$\delta S_z$ and $\bm{\nabla}\cdot\delta\bm{J}^{S_{z}}$ are
the odd functions of $y$.
The disorder strength is $\gamma = 2$ for the Rashba system and  
$\gamma = 1$ (solid) and $\gamma = 5$ (dashed) for the 
Luttinger system, respectively.
As for the Luttinger system,
$\delta S_z$ in $\gamma = 1$
and $\bm{\nabla}\cdot\delta\bm{J}^{S_{z}}$ in $\gamma = 1$ and $5$
are rescaled.
} 
\label{fig:y-profile}
\end{figure}

\begin{figure}[hbt]
\includegraphics[scale=0.3]{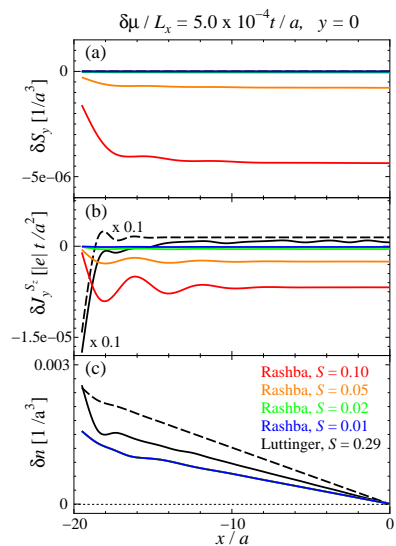}
\caption{
Distribution of (a) $\delta S_y$, (b) $\delta J^{S_z}_y$ 
and (c) $\delta n$ at $y = 0$
in the Rashba and Luttinger systems.
The system size and the disorder strengths are 
the same as in  Fig.~\ref{fig:y-profile}.
$\delta J^{S_z}_y$ of the Luttinger system is rescaled
both in the cases of $\gamma = 1$ and $5$.
Only the region $x < 0$ is shown.
$\delta n$ is the odd functions of $x$.
} 
\label{fig:x-profile}
\end{figure}

We show the obtained results in Figs.1,2, and 3.
Fig.~\ref{fig:dsz-dJs} shows 
the accumulation pattern of  
spin $z$-component $\delta S_{z}$ and the spin current $\delta \bm{J}^{S_z}$
for (a),(b) the Rashba ($S=0.05$) and (c),(d),(e),(f) Luttinger systems.
Here we take the definition of the spin current as
$
\bm{J}^{S_{\mu}}_{\bm{r}\bm{r}'} = \frac{1}{2}
\left(S_{\mu}\bm{J}_{\bm{r}\bm{r}'}+\bm{J}_{\bm{r}\bm{r}'}S_{\mu}\right)
$,
where $\bm{S}$ is the spin-$\frac{1}{2}$ matrices
for the Rashba system and
the spin-$\frac{3}{2}$ matrices for the Luttigner system,
and $\bm{J}_{\bm{r}\bm{r}'}$ is the charge current.
The disorder strength is taken as $\gamma = 2$ for the Rashba system,
and $\gamma = 1$ and $5$ for the Luttinger system.
Then, the inverse lifetime is $1/\tau \sim 0.3t$ for the Rashba system,
$1/\tau \sim 0.25t$ and $1.25t$ for the Luttinger system.
Figure~\ref{fig:y-profile} shows (a) the spin $z$-component $\delta S_{z}$, 
(b) the charge current $\delta J_{x}$ 
and (c) the divergence of the spin current 
$\bm{\nabla}\cdot\delta\bm{J}^{S_{z}}$ at the $x=0$ cross-section, 
while Fig.~\ref{fig:x-profile} 
(a) the spin $y$-component $\delta S_{y}$,
(b) the spin current $\delta J^{S_z}_{y}$
and (c) the electron density $\delta n$
at the $y=0$ cross-section.

As shown in Fig.~\ref{fig:dsz-dJs}~(a), the spin accumulation occurs 
all along the edges for Rashba system. 
This is in sharp contrast to the theoretical 
prediction for the Rashba system in ref.~\cite{Mishchenko}, 
where the spin current and spin accumulation 
is finite only near the electrodes. 
However since the spin current is not the conserved quantity,
not all the spin current contributes to the spin accumulation.
There appears sink ($\bm{\nabla} \cdot \delta \bm{J}^{S_z}< 0$),
which is related to the spin torque density 
and the spin relaxation~\cite{Culcer}, 
near the negative $y$-edge , where the spin current is partly 
absorbed as shown in Fig.~\ref{fig:y-profile}~(c).  
The spin accumulation comes
from the remaining part of $\bm{\nabla} \cdot \delta \bm{J}^{S_z}$
which is not canceled by the torque density and
balancing with the spin relaxation.
Therefore, the bulk SHE,
the direction of the spin current
and the sign of the spin accumulation are consistent.

As for the Luttinger system, 
the spin current is strongly enhanced near the electrodes $x = \pm L_{x}/2$, 
but the resultant spin accumulation is rather flat along the edges
$y = \pm L_{y}/2$.  
The sign of spin accumulation pattern 
for $1/\tau \lesssim t$ ($\gamma=1$)
is opposite to what is expected from that of spin Hall conductivity 
in the bulk
while the direction of spin current in the bulk is consistent.
This is because the spin current near the contact 
gives the opposite contribution 
in this system as shown by the profile of 
$\delta J_{y}^{S_z}$ in Fig.~\ref{fig:x-profile}~(b).
When the relaxation effect is increased, 
the bulk property becomes dominant even in a small system
and the accumulation pattern coincides with 
what is expected from the spin Hall conductivity.
We need more rigorous argument on
the definition of conserved spin current~\cite{PZhang}
in order to investigate this problem furthermore.
As for the Rashba system, as seen in Fig.~\ref{fig:x-profile}~(a), 
the in-plane spin accumulation $\delta S_y$
perpendicular to the electric field
is finite in the bulk as discussed in refs.~\cite{Edelstein,Inoue2}.
On the other hand, 
there appears no in-plane spin accumulation for the Luttigner system
as long as $t^{1,2}_{\bm{r}\bm{r}'} =0$, i.e.,
no hybridization between the decoupled $2\times 2$ matrix spaces
in $4\times 4$ $\Gamma$-matrix space.

\begin{figure}[hbt]
\includegraphics[scale=0.3]{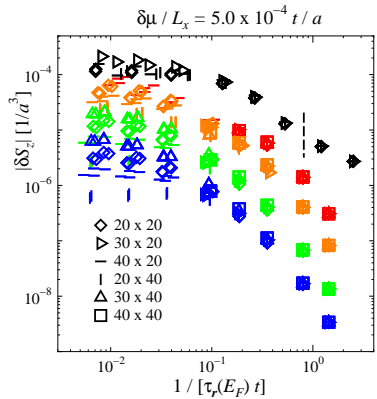}
\caption{
Spin accumulation as a function of the inverse lifetime.
Sample points are the peak values of $|\delta S_z|$ in the region, 
$|x| < 2a$ and $|y|\sim L_y/2$.
The colors are corresponding with those in 
Figs.~\ref{fig:y-profile} and \ref{fig:x-profile}.
The sign of the accumulation pattern in the Luttinger system 
changes around the dashed vertical line.
} 
\label{fig:dsz-scale}
\end{figure}

Next, Fig.~\ref{fig:dsz-scale} shows the inverse lifetime 
dependence of $\delta S_{z}$.
The disorder strength is taken as 
$\gamma = 0.05$, $0.1$, $0.2$, $0.5$, $1$, $2$, $5$ and $10$.
Sample points are the peak values of $|\delta S_z|$ in the region, 
$|x| < 2a$ and $|y|\sim L_y/2$. 
It is seen that the spin accumulation is larger for stronger SOI.
This means that the magnitude of the spin current is the 
more important factor than the spin lifetime.
Therefore, it is concluded that $p$-type GaAs is more advantageous
than $n$-type to observe the spin accumulation due to the ISHE.
Another observation is that the spin accumulation
is rather robust against the relaxation up to $1/\tau \sim 0.1t$.

In order to compare the results with the experiment~\cite{Kato} on 
$n$-type semiconductor,
we fix $t$ and $a$ in the Rashba system as
$t=4$ meV and $a=14$ nm.
The effective mass and the carrier density is estimated as 
$m^{*} \sim 0.05m_{e}$ and $n_\mathrm{3D} \sim 3\times 10^{16}$ cm$^{-3}$,
 respectively.
The Rashba spin splitting is $\Delta_{R} \sim 5.7S$ meV ($S\ll 1$).
The applied electric field in our simulation is $E\sim 0.14$ mV$\mu$m$^{-1}$.
In the case with $\gamma = 2$ which corresponds to $1/\tau \sim 0.3 t$,
The charge resistivity is $\rho_{c}\sim 144$ $\Omega\mu$m,
and the spin Hall resistivity multiplied by $1/e$ is 
$|\rho_{s}| \sim 2.9\times 10^5$, 
$6.4\times 10^4$, $1.2\times 10^4$ and $4.1\times 10^3$ $\Omega\mu$m
for $S = 0.01$, $0.02$, $0.05$ and $0.10$ respectively.
When the results are linearly extrapolated to $E = 10$ mV$\mu$m$^{-1}$,
we obtain :
$|\delta J_{x}|\sim 70$ $\mu$A$\mu$m$^{-2}$,
$|e\delta J^{S_z}_{y}|\sim 35$, $160$, $870$, $2400$ nA$\mu$m$^{-2}$,
and $|\delta S_{z}|\sim 2.6$, $10$, $64$, $150$ $\mu$m$^{-3}$
for  $S = 0.01$, $0.02$, $0.05$ and $0.1$, respectively.
Although the Rashba coupling for $S=0.01$ is still 
an order of magnitude larger than the experimental 
estimation of the Rashba coupling in ref.\cite{Kato}, 
$\rho_{c}$, $\delta J_{x}$ and $\delta S_{z}$
are consistently corresponding to the experimental data.
However, $\rho_{s}$ and $\delta J^{S_z}_{y}$ are not.
This may be attributed to the spin source and sink
where $\bm{\nabla}\cdot\delta \bm{J}^{S_z}\neq 0$.
In the Luttinger system, $t$ and $a$ are fixed as
$t=10$ meV and $a=5$ nm, which correspond to the case with 
$|\xi_{F}|\sim20$ meV and $n_\mathrm{2D} \sim 1.6\times10^{12}$ cm$^{-2}$.
In the case with $\gamma=1$, the charge resistivity is $\rho_{c}\sim 18$ 
$\Omega\mu$m.
In the experiment~\cite{Wunderlich}, 
a current $I_p \sim 100$ $\mu$A is applied
to the $p$-channel of 1.5 $\mu$m width.
Although we cannot exactly read the depth of the channel from the reference,
it is considered to be of the order of 
the length unit in the model, i.e., $a=5$ nm.
In our system, this corresponds to $E_{x} \sim 200$ mV$\mu$m$^{-1}$.
When the results are linearly extrapolated to $E_{x} = 200$ mV$\mu$m$^{-1}$,
$|\delta S_{z}|_\mathrm{2D}\sim 300$ $\mu$m$^{-2}$
and the spin polarization of the holes
is about $2$ $\%$ which is of the order of the observed
circular polarization ($\sim$ 1 $\%$) of light emitted from the LED.

Finally, it is worthwhile to estimate the inverse spin Hall effect
in which a gradient of external magnetic field,
i.e., a spin force, induces a transverse charge current
and/or electric field~\cite{PZhang}.
For the linearly modulated magnetic field,
$\nabla_{y}B_{z}=$ const., applied on an open system,
the induced electric field is given by
$|E_{x}|=|\sigma^{cs}_{xy}\rho_{c}g\mu_{B}\nabla_{y}B_{z}|$, where 
$g$ is the $g$-factor and 
$\sigma^{cs}_{xy}$ is the conductivity of inverse SHE
which has the relation between the spin Hall conductivity 
$\sigma^{sc}_{yx}$ as $\sigma^{cs}_{xy} = -\sigma^{sc}_{yx}$,
via the Onsager's relation.
In the case in which $\nabla_{y}B_{z}=1$ Tcm$^{-1}$, $\rho_{c}=100 \Omega\mu$m
and $1/|e\sigma^{cs}_{xy}|=1000 \Omega\mu$m,
the induced electric field is of the order of 10$\mu$Vcm$^{-1}$,
assuming $g\sim 2$.
Although the above estimation is not large, 
we can enhance this inverse effect
by lightly doped spin Hall insulators,
i.e., materials with low carrier density and large SOI.
This is because larger electric resistivity and spin Hall conductivity
are advantageous for this effect.

In conclusion, we have numerically investigated the spin accumulation 
due to the intrinsic spin Hall effect
in the Rashba and Luttinger systems by using Keldysh formalism.
The distribution of the accumulated spin and the charge and spin currents 
are obtained, which are compared with those of the recent experiments 
obtaining the quantitative agreement.

The authors thank S.~Murakami and  B.~K.~Nikoli\'{c}
for fruitful discussions.
This work is financially supported by NAREGI Grant,
Grant-in-Aids from the Ministry of Education,
Culture, Sports, Science and Technology of Japan.

\end{document}